\documentclass[journal]{IEEEtran}
\usepackage{ifpdf}

\usepackage{cite}

\ifCLASSINFOpdf
\else
\fi
\usepackage[cmex10]{amsmath}
\usepackage{amssymb}
\newtheorem{theorem}{Theorem}
\newtheorem{remark}{Remark}
\newtheorem{definition}{Definition}
\newtheorem{proposition}{Proposition}

\usepackage{algorithmic}

\usepackage{array}

\usepackage{mdwmath}
\usepackage{mdwtab}

\usepackage{eqparbox}
\usepackage{fixltx2e}

\usepackage{stfloats}
\usepackage{url}

\begin{document}
%
\title{A family of sequences with large size and good correlation property arising from $M$-ary Sidelnikov sequences of period $q^d-1$}
%
%
%

\author{Dae San Kim,~\IEEEmembership{Member,~IEEE}
\thanks{}}

%
%

\markboth{Journal of \LaTeX\ Class Files,~Vol.~6, No.~1, January~2007}%
{Shell \MakeLowercase{\textit{et al.}}: Bare Demo of IEEEtran.cls for Journals}
%



\maketitle

\begin{abstract}
Let $q$ be any prime power and let $d$ be a positive integer greater
than 1. In this paper, we construct a family of $M$-ary sequences of
period $q-1$ from a given $M$-ary, with $M|q-1$, Sidelikov sequence
of period $q^d-1$. Under mild restrictions on $d$, we show that the
maximum correlation magnitude of the family is upper bounded by $(2d
-1) \sqrt { q }+1$ and the asymptotic size, as $q\rightarrow
\infty$, of that is $\frac{ (M-1)q^{d-1}}{d }$. This extends the
pioneering work of Yu and Gong for $d=2$ case.
\end{abstract}

\begin{IEEEkeywords}
Correlation, Family size, Sidelnikov sequence, Array structure.
\end{IEEEkeywords}

%
\IEEEpeerreviewmaketitle

\section{Introduction}
%
%
%
%
\IEEEPARstart{I}{n} a code-division multiple-access (CDMA)
communication systems, sequences with low correlation are required
for synchronization and minimization of multiple-access
interference. For adaptive modulation schemes, sequences with
variable lengths and alphabet sizes are desirable to maximize data
rate according to channel characteristics. Moreover, a large number
of distinct sequences are needed to support as many users as
possible.

In \cite{ref06}, for any prime power $q$ and a positive integer $M$,
with $M|q-1$, Sidelnikov introduced $M$-ary sequences(called
Sidelnikov sequences) of period $q-1$, which  have the maximum
out-of-phase autocorrelation magnitude of 4. Kim and Song in
\cite{ref03} showed that the cross-correlation of an $M$-ary
Sidelnikov sequence of period $q-1$ and its constant multiple has
the maximum magnitude of $\sqrt{q}+3$.

Sidelnikov sequences can be used in constructing a large number of
distinct sequences. In this direction of efforts, one refers to the
papers \cite{ref02}, \cite{ref04}, and \cite{ref08}-\cite{ref11}.

In this paper, we consider $M$-ary Sidelnikov sequences, with
$M|q-1$, of period $q^d-1$($q=p^n$ a prime power) and study the
$(q-1)\times (\frac{q^d-1}{q-1})$ array structure of such sequences.
Then we construct a family of $M$-ary sequences with period $q-1$,
with large size and good correlation property. It is formed as the
constant multiples of those column sequences corresponding to a set
of $q$-cyclotomic coset representatives mod $(\frac{q^d-1}{q-1})$.
Under the mild restrictions on $d$(cf. (9)), it is shown that the
maximum correlation magnitude of the family is upper bounded by
$(2d-1)\sqrt{q}+1$, and the asymptotic size, as $q\rightarrow
\infty$, of that is $\frac{(M-1)q^{d-1}}{d}$. Also, we derive an
exact but less explicit expression of the size of the family of
sequences by using a result of Yucas \cite{ref12}. One refers to the
tables either in \cite{ref02} or \cite{ref09} to compare our result
with the known ones. This generalizes the pioneering work of Yu and
Gong for $d=2$ case in \cite{ref09} and \cite{ref10}.

\section{Preliminaries}

We will use the following notations throughout this paper.

\begin{itemize}
\item $p$ a prime number,
\item $n$ a positive integer,
\item $q=p^n$,
\item $\mathbb{F}_q$ the finite field with $q$ elements,
\item $\mathbb{F}_{q^d}$ the finite field with $q^d$ elements, with $d\geq2$,
\item $M$ a positive divisor of $q-1$, with $M\geq2$,
\item $w_M = \exp(\frac{2\pi i}{M})$,
\item $\alpha$ a fixed primitive element of $F_{q^d}$,
\item $\beta=\alpha^{\frac{q^d-1}{q-1}}=N(\alpha)$ a primitive element
of $\mathbb{F}_q$,
\item $N$ the norm map from $\mathbb{F}_{q^d} \rightarrow \mathbb{F}_q$, given by
$N(x)=x^{\frac{q^d-1}{q-1}}$,
\item $Tr$ the trace map from $\mathbb{F}_{q^d} \rightarrow \mathbb{F}_q$, given by $Tr(x) =
\sum_{j=0}^{d-1}x^{q^j}$,
\item $\psi$ the multiplicative character of $\mathbb{F}_q$ of order $M$, defined by $\psi(x) = \exp(\frac{2\pi i \log_\beta x}{M})= w_M^{\log_\beta
x}$.
\end{itemize}

Here we recall that, for any fixed primitive element $\beta$ of
$\mathbb{F}_q$, a logarithm over $\mathbb{F}_q$ is defined by
\begin{equation*}
\log_\beta x =
\left\{
  \begin{array}{ll}
    t, & \hbox{if $x = \beta^t$ ($0 \leq t \leq q-2$),} \\
    0, & \hbox{if $x=0$.}
  \end{array}
\right.
\end{equation*}

\noindent so that, in particular, $\psi(0)=1$. This convention is
not the usual one requiring $\psi(0)=0$. However, this agreement
turns out to be very convenient, as this has been fruitfully
demonstrated in the papers \cite{ref08}-\cite{ref11}.

Again, for any fixed primitive element $\beta$ of $\mathbb{F}_q$,
the $M$-ary Sidelnikov sequence $s(t)$ of period $q-1$ is defined as
\begin{equation}\label{equ01}
s(t) =
\left\{
  \begin{array}{ll}
    k, & \hbox{if $\beta^t \in D_k$,} \\
    0, & \hbox{if $\beta^t = -1$,}
  \end{array}
\right.
\end{equation}
where $D_{k} = \{ \beta^{Mj+k}-1|0 \leq j < \frac{q-1}{M} \}$, for
$0\leq k\leq M-1$.

\noindent It is clear that $s(t)$ can be defined equivalently as
\begin{equation}\label{equ02}
s(t) \equiv \log_\beta(\beta^t+1) \mod M,
\end{equation}
or as
\begin{equation*}
w_M^{s(t)} = \psi(\beta^t + 1).
\end{equation*}

The Weil's estimate for multiplicative character sums is well
known(cf. \cite{ref05}, Theorem 5.41). In [7, Corollary 2.3], Wan
generalized his estimate to the case of multiple multiplicative
character sums. On the other hand, Yu and Gong(cf.
\cite{ref08}-\cite{ref11})introduced a refined version of Wan's
bound that works under the assumption that the value of the
multiplicative characters at 0 are equal to 1 rather than the
traditional 0. Here we state only a special case that is just
suitable for our purpose.

\vspace{2mm}

\begin{theorem}[\cite{ref07}, \cite{ref09}]\label{thm01}
Let $f_1 (x), \ldots, f_m (x)$ be monic distinct irreducible
polynomials over $\mathbb{F}_q$ with degrees $d_1, \ldots, d_m$,
with $e_j$ the number of distinct roots in $\mathbb{F}_q$ of
$f_j(x)$($j=1,\ldots,m$). Let $\psi_1,\ldots,\psi_m$ be nontrivial
multiplicative characters of $\mathbb{F}_q$, with $\psi_j (0)=1$
($j=1,\ldots,m$). Then, for $a_1, \ldots,a_m \in
\mathbb{F}_q^{\times}$, we have the estimate
\begin{equation}\label{equ03}
\begin{split}
& \left|\sum_{ x \in \mathbb{F}_q }  \psi_1 ( a_1 f_1 ( x)) \cdots
\psi_m (a_m f_m
(x)) \right| \\
& \quad \leq \left(\sum_{ j=1}^{ m} d_j -1\right) \sqrt { q}+
\sum_{j=1 }^{m}e_j.
\end{split}
\end{equation}
\end{theorem}

\section{Array structure of the $M$-ary Sidelnikov sequences of
period $q^d-1$}

Here we investigate the $(q-1)\times(\frac{q^d-1}{q-1})$ array
structure of $M$-ary Sidelnikov sequences of period $q^d-1$, with
$M|q-1$. This is a generalization of the $d=2$ case in \cite{ref09}
and \cite{ref10} that has its origin in the paper \cite{ref01}.

\vspace{2mm}

\begin{theorem}\label{thm02}
Let $\{s(t)\}$ be an $M$-ary Sidelnikov sequences of period $q^d-1$,
with $M|q-1$. Then
\begin{equation}\label{equ04}
s(t) \equiv \log_{\beta} (N(\alpha^t +1)) \mod M,
\end{equation}
where $0\leq t \leq q^d-2$.

In other words,
\begin{equation*}
s(t) = \left\{
         \begin{array}{ll}
           0, & \hbox{if $N(\alpha^t+1)=0$,} \\
           k, & \hbox{if $N(\alpha^t + 1)\in S_k$,}
         \end{array}
       \right.
\end{equation*}
where $S_k = \{ \beta^{Mj+k} | 0 \leq j < \frac{q-1}{M}\}$, for
$0\leq k \leq M-1$.
\end{theorem}

\vspace{2mm}

\begin{remark}\label{rem03}
Note here that the sets $S_k$ are different from those $D_k$ in
(\ref{equ01}).

\vspace{2mm}

\begin{proof}
By definition of Sidelnikov sequence,
\begin{equation*}
s(t) \equiv y(t) \mod M, \text{ with } y(t) =
\log_\alpha(\alpha^t+1).
\end{equation*}
To prove the statement, we may assume that $N(\alpha^t+1)\not=0$.
Then, with $N(\alpha^t +1)=\beta^{x(t)}$,
\begin{equation*}
\begin{split}
\frac{q^d-1}{q-1} y(t) & \equiv \log_{\alpha} (\alpha^t +1)^{\frac{
q^d -1}{q-1}} \\
& \equiv \log_{\alpha} N(\alpha^t +1)\\
& \equiv \log_\alpha \alpha^{\frac{ q^d -1}{ q-1} x(t)}\\
& \equiv \frac{q^d -1}{ q-1}x(t) \mod q^d-1.
\end{split}
\end{equation*}
This implies that
\begin{equation*}
x(t) \equiv y(t) \mod q-1,
\end{equation*}
and hence that, as $M|q-1$,
\begin{equation*}
x(t) \equiv y(t) \mod M,
\end{equation*}
Thus
\begin{equation*}
s(t) \equiv y(t) \equiv x(t) \equiv \log_\beta N(\alpha^t+1) \mod M.
\end{equation*}
\end{proof}
\end{remark}

We list the sequence $\{s(t)\}(0\leq t\leq q^d-2)$ as an
$(q-1)\times (\frac{q^d-1}{q-1})$ array so that the $l$-th column
$v_l(t)(0\leq t\leq q-2)$ of the array is given by:

\begin{equation*}
v_l (t) = s\left(\left(\frac{q^d -1}{q-1}\right) t+l\right), \quad
(0 \leq l \leq \frac{q^d -1}{q-1}-1 ).
\end{equation*}
Then
\begin{equation}\label{equ05}
v_l (t) \equiv \log_{\beta} ( N(\alpha^l  \beta^t + 1)) \mod M.
\end{equation}

Let $f_l(x)$ be the polynomial of degree $d$ over $\mathbb{F}_q$
given by: for any nonnegative integer $l$,
\begin{equation*}
\begin{split}
f_l (x) & = N(\alpha^l x + 1)\\
& = (\alpha^l x + 1)(\alpha^{lq}x + 1) \cdots (\alpha^{l q^{d-1}}x
+1)\\
& = \beta^l x^d + \cdots + Tr(\alpha^l ) x + 1.
\end{split}
\end{equation*}

Then
\begin{equation}\label{equ06}
v_l (t) \equiv \log_{\beta} f_l (\beta^t ) \mod M.
\end{equation}

For each $l (0 \leq l \leq \frac{q^d -1}{q-1 }-1 )$,
\begin{equation}\label{equ07}
\begin{split}
f_l (x) &= \beta^l N (x + \alpha^{-l} )\\
& = \beta^l (x + \alpha^{-l})(x + \alpha^{-lq}) \cdots(x +
\alpha^{-lq^{d-1}})\\
& = \beta^l p_l (x)^{ \frac{d}{ d_l}},
\end{split}
\end{equation}
where $p_l(x)$ is the irreducible polynomial over $\mathbb{F}_q$ of
$-\alpha^{-l}$ of degree $d_l$. Note here that $d_l|d$.

\vspace{2mm}

\begin{remark}\label{rem04}
Note that the $q$-cyclotomic coset containing $l(0\leq l \leq q^d-2)
\mod q^d-1$ is
\begin{equation*}
C_l = \{ l, ql, \cdots, q^{d_l -1} l \},
\end{equation*}

\noindent where each $q^j l$ is reduced modulo  $q^d-1$, $d_l$ is
the smallest positive integer satisfying $q^{d_l} l\equiv l \mod q^d
-1$, and
\begin{equation}\label{equ08}
p_l (x) = \prod_{j \in C_l} (x + \alpha^{-j} ).
\end{equation}
Here $l$ is taken as the smallest positive integer in $C_l$ modulo
$q^d-1$, as usual.
\end{remark}

\vspace{2mm}

\begin{proposition}\label{pro05}
\begin{enumerate}
\item $v_l (t)=v_{lq} (t)$.
\item $p_l (x)$ has no roots in $\mathbb{F}_q$, for $l$, with $1 \leq l \leq \frac{q^d -1}{q-1
}-1$.
\item For nonnegative integers $l_1,l_2$, with $l_1\equiv l_2 \mod \frac{q^d-1}{q-1}$, $v_{l_1}(t)$ and $v_{l_2}(t)$ are cyclically equivalent.
\item $v_{\frac{q^d -1}{q-1} - \frac{q^{d-1}-1}{q-1}l} (t) \equiv v_{l} (t-l+1) \mod
M$, so that $v_{ \frac{q^d-1}{q-1}-\frac{q^{d-1}-1}{q-1 }l} (t)$ and
$v_l(t)$ are cyclically equivalent for each $l(1\leq l \leq q)$.
\end{enumerate}

\vspace{2mm}

\begin{proof}
\begin{enumerate}
\item $f_l (x) = f_{lq} (x)$, so that $v_l(t)=v_{lq}(t)$, by
(\ref{equ06}).
\item This follows from the observation that $d_l =  1$ iff $\alpha^l \in \mathbb{F}_q$
iff $\frac{q^d-1}{q-1}|l$.
\item is easy to see.
\item This is a generalization of the result for $d=2$ discovered by Yu and
Gong in \cite{ref09} and \cite{ref10}: for each $l(1\leq l \leq q)$,
\begin{equation*}
\begin{split}
v_{\frac{q^{d}-1}{q-1} - \frac{q^{d-1}-1}{q-1} l} (t) & \equiv
\log_{\beta} ( N(\alpha^{-\frac{q^{d-1} -1}{q-1}l}
\beta^{t+1}+1))\\
& \equiv \log_{\beta} ( N(\alpha^{-\frac{q^{d-1}-1}{q-1} ql}
\beta^{t+1}+1))\\
& \equiv \log_{\beta} ( N(\alpha^{- \frac{q^{d-1}-1}{q-1} ql-l+l}
\beta^{t+1}+1))\\
& \equiv \log_{\beta} ( N(\alpha^l \beta^{t-l+1} +1))\\
& \equiv v_{l}(t-l+1)\mod M.
\end{split}
\end{equation*}
Also, this follows from 1) and 3), since
\begin{equation*}
q \left( \frac{q^d-1}{q-1}- \frac{q^{d-1}-1}{q-1}l \right) \equiv l
\mod \frac{q^d-1}{q-1}.
\end{equation*}
\end{enumerate}
\end{proof}
\end{proposition}

\vspace{2mm}

\begin{remark}\label{rem06}
Because of 1) and 3) of Proposition \ref{pro05}, we are led to
consider the $q$-cyclotomic cosets $\mod \frac{q^d-1}{q-1}$. Recall
that the $q$-cyclotomic coset containing $l (0\leq l \leq
\frac{q^d-1}{q-1} - 1) \mod \frac{q^d-1}{q-1}$ is
\begin{equation*}
\hat{C_l} = \{ l,ql,\ldots,q^{m_l -1} l \},
\end{equation*}
where each $q^jl$ is reduced modulo $\frac{q^d-1}{q-1}$, $m_l$ is
the smallest positive integer satisfying $q^{m_l}l\equiv l \mod
\frac{q^d-1}{q-1}$. Again, here $l$ is taken as the smallest
positive integer in $\hat{C_l}$ modulo $\frac{q^d-1}{q-1}$, as
usual. Here $m_l | d_l$. So if $q_l (x) = \prod_{j \in \hat{C_l}}
(x+\alpha^{-j})$, then
\begin{equation*}
p_l (x) = q_l (x) q_l (x)^{\sigma^{m_l} }q_l (x)^{\sigma^{2 m_l} }
\cdots q_l(x)^{\sigma^{d_l -m_l}}.
\end{equation*}
Here $\sigma$ is the Frobenius automorphism of $\mathbb{F}_{q^d}$
over $\mathbb{F}_q$, given by $\sigma(a)=a^q$, so that
\begin{equation*}
q_l (x)^{\sigma^{i m_l}} =  \prod_{j \in \hat{C_l}}(x+\alpha^{-j
q^{i m_l}})\quad (0 \leq i \leq \frac{d_l}{ m_l}-1 ).
\end{equation*}
\end{remark}

\section{Construction of a family of sequences}

Here we construct a family $\Sigma$ of $M$-ary sequences with period
$q-1$, consisting of the constant multiples of those column
sequences $v_l(t)$ corresponding to a set of $q$-cyclotomic coset
representatives $\mod \frac{q^d-1}{q-1}$, for the set consisting of
$l(1\leq l\leq \frac{q^d-1}{q-1})$. Then it is shown that, under
mild restrictions on $d$ (cf. (\ref{equ09}), it has a large family
size and good correlation property. Actually, we show that the
maximum correlation magnitude of the family is upper bounded by
$(2d-1)\sqrt{q}+1$, and the asymptotic size, as $q\rightarrow
\infty$, of that is $\frac{(M-1)q^{d-1}}{d}$. Also, we derive an
exact but less explicit expression of the size of the family of
sequences by using a result of Yucas(cf. Theorem \ref{thm15}). This
generalizes the pioneering work of Yu and Gong for $d=2$ case in
\cite{ref09} and \cite{ref10}.

\vspace{2mm}

\begin{definition}\label{def07}
Let $\Lambda$ be the set of all integers $l(0 \leq l \leq \frac{q^d
-1}{ q-1}-1)$ consisting of the smallest $q$-cyclotomic coset
representative from each $q$-cyclotomic coset $\mod
\frac{q^d-1}{q-1}$.
\end{definition}

\vspace{2mm}

\begin{proposition}\label{pro08}
1) $|\Lambda|$ = the number of $q$-cyclotomic cosets $\mod
\frac{q^d-1}{q-1}$ = the number of monic irreducible factors of
$x^{\frac{q^d-1}{q-1}}-1$.

2) Let $p(x) = x^e + \cdots + (-1)^{e} b$ be a monic irreducible
factor of $x^{ \frac{q^d -1}{q-1}}-1$. Then $e|d$, and
$b^{\frac{d}{e}}=1$.

\vspace{2mm}

\begin{proof}
1) The first equality is just Definition \ref{def07}. Let $\gamma =
\alpha^{q-1}$ be a primitive $(\frac{q^d-1}{q-1})$-th root of unity
in $\mathbb{F}_{q^d}$. Then, with $M^{(l)}(x) = \prod_{j \in
\hat{C_l}}(x- \gamma^j)$ denoting the irreducible polynomial of
$\gamma^l$ over $\mathbb{F}_q$, we have
\begin{equation*}
x^{ \frac{q^d -1}{q-1}}-1 = \prod_{l \in \Lambda } M^{(l)} (x).
\end{equation*}
Thus we have the desired equality.

2) Clearly, $e|d$. For a root $\alpha$ of $p(x)$ in
$\mathbb{F}_{q^d}$, $N(\alpha)=1$, and $((-1)^e
b)^{\frac{d}{e}}=(-1)^{d } b^{ \frac{d}{e}}$ is the constant term of
$p(x)^{ \frac{d}{e}}=x^d+\cdots +(-1)^d N(\alpha)$.
\end{proof}
\end{proposition}

\vspace{2mm}

Assume from now on that
\begin{equation}\label{equ09}
(d,q-1)=1, \quad d < \frac{ \sqrt{q} - \frac{2}{\sqrt{q}}+1}{2}.
\end{equation}

\vspace{2mm}

\begin{proposition}\label{pro09}
Let $l_1,l_2$ be elements in $\Lambda\setminus\{0\}$, and let $\tau
(0\leq \tau \leq q-2)$ be an integer. Then $p_{l_1}(x)$ and $\beta^{
-\tau d_{l_2}} p_{l_2} (\beta^\tau x)$ are distinct irreducible
polynomials over $\mathbb{F}_q$, unless $l_1=l_2$ and $\tau=0$. Here
\begin{equation}\label{equ10}
\begin{split}
& \beta^{-\tau d_{l_2}}p_{l_2} (\beta^\tau x ) \\
& = (x+\alpha^{-l_2} \beta^{-\tau} )(x+\alpha^{-l_2 q}
\beta^{-\tau})\cdots (x+\alpha^{-l_2 q^{d_{l_2}-1}} \beta^{-\tau}).
\end{split}
\end{equation}

\vspace{2mm}

\begin{proof}
We know that $p_{l_1} (x)$ and $\beta^{ -\tau  d_{l_2}} p_{l_2}
(\beta^\tau x )$ are irreducible polynomials over $\mathbb{F}_q$.
Assume that they are the same. Then $\alpha^{-l_1} = \alpha^{- l_2
q^s} \beta^{-\tau}$, for some nonnegative integer $s(0 \leq s \leq
d_{l_2}-1)$, and hence $l_1 \equiv l_2 q^s + \tau ( \frac{q^d-1}{
q-1}) \mod q^d-1$. So $l_1\equiv l_2 q^s \mod \frac{q^d-1}{q-1}$ and
thus $l_1$ and $l_2$ are in the same $q$-cyclotomic coset $\mod
\frac{q^d-1}{q-1}$. This implies $l_1=l_2$. Now, $l_1 \equiv l_1 q^s
\mod \frac{q^d-1}{q-1}$, and hence $l_1 (q^{s}-1)=\tau' ( \frac{q^d
-1}{ q-1})$.

Observe that we have $\frac{q^{d} -1}{q-1}= f(q) (q-1)+d$, for $f
(q)= \sum_{ j=1}^{ d-1}jq^{d-j-1}$, and hence that $(q-1, \frac{q^d
-1}{q-1})=(q-1,d)=1$. Hence $l_1 (q^s -1)\equiv 0 \mod q^d-1$, and
so $d_{l_1}|s$. As $0\leq s \leq d_{l_1}-1=d_{l_2} -1$, we have
$s=0$. In all, $l_1\equiv l_1+ \tau ( \frac{q^d -1}{ q-1})\mod q^d-1
$ which implies $q-1|\tau$, and therefore $\tau=0$.
\end{proof}
\end{proposition}

\vspace{2mm}

\begin{definition}\label{def10}
Let $\Sigma$ be the family consisting of $M$-ary sequences of period
$q-1$, given by
\begin{equation*}
\Sigma =\{ c v_l (t) |1 \leq c \leq M-1,l \in \Lambda
 \setminus  \{0 \} \}.
\end{equation*}
\end{definition}

\vspace{2mm}

\begin{remark}\label{rem11}
When $d=2, \Lambda \setminus \{0 \}= \{1,\ldots,[\frac{q+1}{ 2}]\}$.
This follows from the simple observation that the $q$-cyclotomic
coset containing $l \mod q+1$ is $\hat{C_l}=\{ l,ql\} $, and
$ql\equiv q-l +1\mod q+1$. So the family $\textbf{S}_v$ considered
in \cite{ref09} and \cite{ref10} is identical to our $\Sigma$, for
$q=p^n$ even and contains $M-1$ less sequences, namely $c v_{
\frac{q+1}{2}} (t)(1\leq c \leq M-1 )$, for $q$ odd.
\end{remark}

\vspace{2mm}

Recall that the maximum correlation of $\Sigma$, $\delta_{\max} =
\delta_{\max}(\Sigma)$, is defined as the maximum absolute value of
all nontrivial auto- and cross-correlations of the sequences in
$\Sigma$.

\vspace{2mm}

\begin{theorem}\label{thm12}
For the family $\Sigma =\{ c v_l (t) |1\leq c \leq M-1,l \in \Lambda
\setminus \{0 \} \}$ of $M$-ary sequences of period $q-1$, we have
\begin{equation*}
\delta_{\max} (\Sigma) \leq (2d-1) \sqrt{q}+1.
\end{equation*}
\begin{proof}
Assume that $l_1 \not= l_2(l_1,l_2 \in \Lambda \setminus \{0 \} )$
or $\tau$ is in the range $1 \leq \tau \leq q-2$. Then $p_{l_1} (x)$
and $\beta^{ -\tau  d_{l_2}} p_{l_2} (\beta^\tau x )$ are distinct
irreducible polynomials over $\mathbb{F}_q$, by Proposition
\ref{pro09}. The cross-correlation function $R(\tau) =
R_{c_1,l_1,c_2,l_2}(\tau)$ between the sequence $c_1 v_{l_1} (t)$
and $c_2 v_{l_2} (t)$ in $\Sigma$ is given by
\begin{equation}\label{equ11}
\begin{split}
R(\tau) & = \sum_{t=0}^{q-2} {w_M}^{c_1 v_{l_1} (t) - c_2 v_{l_2}
(t + \tau)}\\
& = \sum_{t=0}^{q-2} \psi^{c_1} (f_{l_1} (\beta^{t} ) ) \psi^{M-c_2} (f_{l_2} (\beta^{t + \tau} ))\\
& = \sum_{x \in \mathbb{F}_{q}} \psi_{1} (p_{l_{1}} (x)) \psi_{2}
(\beta ^{- \tau d_{l_{2}}} \times \beta^{\tau d_{l_{2}}} p_{l_{2}} (
\beta ^{\tau } x))-1,
\end{split}
\end{equation}
where $\psi_1 = \psi^{c_1 \frac{d}{d_{l_1}}}$ and $\psi_2 =
\psi^{c_2 \frac{d}{d_{l_2}}}$. Observe that both $c_1 \frac{d}{
d_{l_1}}$ and $c_2 \frac{d}{ d_{l_2}}$ are not divisible by $M$ and
hence $\psi_1$ and $\psi_2$ are both nontrivial, since $(d,q-1)=1$.
In view of (\ref{equ03}), the sum in (\ref{equ11}) in absolute value
is
\begin{equation*}
\begin{split}
| \sum_{x \in \mathbb{F}_q} & \psi_{1} (p_{l_1} (x)) \psi_{2} (
\beta^{-\tau d_{l_2}} \times \beta^{\tau d_{l_2}} p_{l_2} (
\beta^{\tau} x)) | \\
& \leq (d_{l_{1}} +d_{l_{2}} -1) \sqrt{q}\\
& \leq (2d-1) \sqrt{q}.
\end{split}
\end{equation*}
So we get the desired result in this case. Note here that $p_{l_1}
(x)$ and $\beta^{ -\tau  d_{l_2}} p_{l_2} (\beta^\tau x )$ have no
roots in $\mathbb{F}_q$, by Proposition \ref{pro05} 2), and
(\ref{equ10}). Then we consider the case that $c_1\not= c_2$, but
$l_1=l_2$ and $\tau=0$. In this case,
\begin{equation*}
\begin{split}
R(\tau) & = \sum_{t=0}^{q-2} w_M^{(c_1 -c_2 ) v_{l_1} (t) }\\
& = \sum_{x \in \mathbb{F}_{q}} \psi^{*} (p_{l_{1}} (x))-1,
\end{split}
\end{equation*}
where $\psi^* = \psi^{(c_1 -c_2 ) \frac{d}{d_{l_1}}}$ is nontrivial,
as $(c_1 - c_2 ) \frac{d}{d_{l_1}}$ is not divisible by $M$. So, by
the classical Weil's theorem(the $m=1$ case of Theorem \ref{thm01}),
\begin{equation*}
\begin{split}
| R(\tau) | & \leq (d_{l_1} -1) \sqrt{q}+1 \\
& \leq (d-1) \sqrt{q}+1.
\end{split}
\end{equation*}
Note that these take care of the cases that $(c_1,l_1 ) \not=
(c_2,l_2 )$ and $(c_1,l_1 ) = (c_2,l_2 )$, but with $\tau\not=0$.
\end{proof}
\end{theorem}

\vspace{2mm}

\begin{theorem}\label{thm13}
The sequences in the family $\Sigma = \{ c v_l (t) |1 \leq c \leq
M-1, l \in \Lambda \setminus \{0 \} \}$ are cyclically inequivalent.

\vspace{2mm}

\begin{proof}
If $c_1 v_{l_1} (t)$ and $c_2 v_{l_2} (t)$ are cyclically
equivalent, then, for some $\tau (0 \leq \tau \leq q-2)$, $c_1
v_{l_1} (t) = c_2 v_{l_2} (t + \tau)$ and hence
\begin{equation*}
\begin{split}
q-1 & = \sum_{t=0}^{q-2} w_M^{c_1 v_{l_1} (t) - c_2 v_{l_2}
(t + \tau)}\\
& = | \sum_{t=0}^{q-2} w_M^{c_1 v_{l_1} (t) - c_2 v_{l_2}
(t + \tau)} |\\
& \leq | \sum_{x \in \mathbb{F}_{q}} \psi_{1} (p_{l_{1}} (x))
\psi_{2} (
\beta^{- \tau d_{l_{2}}} \times \beta^{\tau d_{l_{2}}} p_{l_{2}} ( \beta^{\tau} x)) | + 1\\
& \leq (2d-1) \sqrt{q}+1,
\end{split}
\end{equation*}
if $(c_1,l_1 ) \not= (c_2,l_2)$. Here $\psi_1 = \psi^{c_1 \frac{d}
{d_{l_1}}}$ and $\psi_{2} = \psi^{c_{2} \frac{d}{d_{l_{2}}}}$. This
is impossible in view of our assumption in (\ref{equ09}). Thus $c_1
v_{l_1} (t)$ and $c_2 v_{l_2} (t)$ are the same.
\end{proof}
\end{theorem}

\vspace{2mm}

\begin{remark}\label{rem14}
Under the mild restrictions in (\ref{equ09}), we proved Proposition
\ref{pro09}, and Theorems \ref{thm12} and \ref{thm13}.  Assume that
$d=2$. The second condition in (\ref{equ09}) needed in proving
Theorem \ref{thm13} misses only a few values of $q$. Namely,
$q=2,4,8,3,9,5,7$, and 11. Note that $(2,q-1)=1$ for $q$ even and
$(2,q-1)=2$ for $q$ odd. Suppose we are in the latter case. Then the
first condition in (\ref{equ09}) is not necessary in showing
Theorems \ref{thm12} and \ref{thm13}, since $\frac{d}{d_{l_1}}=
\frac{d}{d_{l_2}}=1$, and so the $\psi_1$ and $\psi_2$ are
nontrivial. In addition, if we replace $\Lambda\setminus\{0\}$ by
$\Lambda\setminus \{ 0,\frac{q+1}{2} \}= \{1,\ldots,\frac{q-1}{
2}\}$, then one easily checks that the statement of Proposition
\ref{pro09} holds true.
\end{remark}

\vspace{2mm}

\begin{theorem}[12, Theorem 3.5]\label{thm15}
Let $A_f = \{ r | ~r|q^f -1 \text{ but } r \text{ does not divide }
q^g-1 \text{ for } 1\leq g < f\}$, for each positive integer $f$,
and, for $r\in A_f$, write $r=d_{rf}m_{rf}$, with
$d_{rf}=(r,\frac{q^f -1}{q-1})$.

Assume $b \in \mathbb{F}_q^\times$ has order $m$, and let $N(f,b,q)$
denote the number of monic irreducible polynomials over
$\mathbb{F}_q$ of degree $f$ with constant term $(-1)^f b$. Then
\begin{equation}\label{equ12}
N(f,b,q) = \frac{1}{f \phi(m)} \sum_{ \substack{r \in A_f  \\ m_{r
f}= m} } \phi (r).
\end{equation}
\end{theorem}

\vspace{2mm}

\begin{theorem}\label{thm16}
The size of the family $\Sigma = \{ c v_l (t) |1 \leq c \leq M-1,l
\in \Lambda \setminus \{0 \} \}$, with the notations in the above,
can be expressed as:
\begin{equation}\label{equ13}
|\Sigma | = (M-1) ( |\Lambda |-1),
\end{equation}
where the number of monic irreducible factors $|\Lambda|$ of $x^{
\frac{q^d -1}{q-1 }} -1$  is given by
\begin{equation}\label{equ14}
\sum_{ e |d } \frac{1}{e} \sum_{ m | \frac{d}{e} } \sum_{\substack{r
\in A_e \\ m_{re} = m }} \phi(r).
\end{equation}

\vspace{2mm}

\begin{proof}
Clearly, we have (\ref{equ13}). By Proposition \ref{pro08} 1), the
size of $\Sigma$ is also given by
\begin{equation*}
\begin{split}
& | \Sigma | =  (M-1) \times \text{((the number of monic irreducible} \\
& \qquad \qquad \qquad \qquad \text{ factors of $x^{ \frac{q^d
-1}{q-1}}-1$)$-1$)}.
\end{split}
\end{equation*}
Thus we only need to verify that the number of irreducible factors
$|\Lambda|$ of $x^{ \frac{q^d-1}{q-1}}-1$ is given by the expression
in (\ref{equ14}). In view of Proposition \ref{pro08} 2), that number
is equal to
\begin{equation}\label{equ15}
\begin{split}
& \sum_{ e | d} \sum_{ b^{ \frac{d}{e}}=1} \text{(\# of monic irreducible factors over $\mathbb{F}_q$ of $x^{ \frac{q^d-1}{q-1}}-1$,}\\
& \qquad \text{ with degree $e$ and the constant term equal to
$(-1)^eb$)}\\
& = \sum_{ e | d} \sum_{ m | \frac{d}{e}} \sum_{ \substack{b \\
o(b)=m}} \text{(\# of monic irreducible polynomials over $\mathbb{F}_q$}\\
& \qquad \text{ with degree $e$ and the constant term equal to
$(-1)^eb$)}\\
& = \sum_{ e | d} \sum_{ m | \frac{d}{e} } \sum_{ \substack{b \\o(b)
= m}} N(e,b,q).
\end{split}
\end{equation}
The desired result now follows from (\ref{equ12}).
\end{proof}
\end{theorem}

\vspace{2mm}

\begin{remark}\label{rem17}
Let's consider the case of $d=2$. In that case,
\begin{equation*}
\begin{split}
|\Lambda| & = \sum_{\substack{r \in A_1 \\ m_{r 1}=1 }} \phi(r)+\sum
_{\substack{r\in A_1 \\ m_{r1}=2} } \phi(r) + \frac{1}{2} \sum_{
\substack{r \in A_2 \\ m_{r2}=1}} \phi(r) \\
& = 1 + \sum_{2|q-1} 1+ \frac{1}{2} \sum_{\substack{r|q+1 \\ r
\not=1, 2}}\phi (r)
\end{split}
\end{equation*}
and hence
\begin{equation}\label{equ16}
| \Lambda | - 1 = \left[\frac{q+1}{2}\right] = \left\{
                                    \begin{array}{ll}
                                      \frac{q+1}{2}, & \hbox{if $q$ odd,} \\
                                      \frac{q}{2}, & \hbox{if $q$ even.}
                                    \end{array}
                                  \right.
\end{equation}
This is what is expected(cf. Remark \ref{rem11}).
\end{remark}

\vspace{2mm}

 The next theorem follows from [7, Theorem 5.1] by
taking $f(T)=T$. It gives an estimate for $N(f,b,q)$ in
(\ref{equ12}).

\vspace{2mm}

\begin{theorem}[\cite{ref07}]\label{thm18}
Let $N(f,b,q)$ denote the number of monic irreducible polynomials
over $\mathbb{F}_q$ of degree $f$ with constant term $(-1)^fb$, for
some element $b\in \mathbb{F}_q^\times$. Then
\begin{equation}\label{equ17}
\left|N(f,b,q) - \frac{q^{f}}{f(q-1)} \right| \leq \frac{2}{f} q^{
\frac{f}{2}}.
\end{equation}
\end{theorem}

\vspace{2mm}

\begin{theorem}\label{thm19}
The asymptotic size of $\Sigma = \{ c v_l (t)  |1 \leq c \leq M-1,l
\in  \Lambda \setminus \{0 \} \}$, as $q\rightarrow \infty$, is
given by:
\begin{equation*}
|\Sigma| \sim \frac{(M-1) q^{d-1}}{ d }, \text{ as }q\rightarrow
\infty.
\end{equation*}
\begin{proof}
Assume first that $d>2$. From (\ref{equ15}) and (\ref{equ17}),
\begin{equation*}
\left| | \Lambda |- d \sum_{ e |d }  \frac{q^e}{e^2 (q-1) } \right|
\leq 2d \sum_{e|d }  \frac{q^{e/2}}{e^2}.
\end{equation*}
This implies that
\begin{equation*}
|\Lambda| \sim \frac{q^{d -1}}{d}, \text{ as }q\rightarrow \infty,
\end{equation*}
and hence
\begin{equation}\label{equ18}
|\Sigma| \sim \frac{(M-1)q^{d-1}}{d}, \text{ as }q\rightarrow
\infty.
\end{equation}
Even for $d=2$, we get the same result as in (\ref{equ18}). Indeed,
from (\ref{equ16}), we have
\begin{equation*}
|\Sigma |=(M-1)[ \frac{q+1}{2}] \sim \frac{(M-1) q}{2}, \text{ as
}q\rightarrow \infty.
\end{equation*}
\end{proof}
\end{theorem}

\section{Conclusion}

In this paper, starting with $M$-ary Sidelnikov sequences, with
$M|q-1$, of period $q^d-1$($q=p^n$ a prime power) and considering
the $(q-1)\times (\frac{q^d-1}{q-1})$ array structure of such
sequences, we constructed a family of $M$-ary sequences with period
$q-1$, with large size and good correlation property. It is formed
as the constant multiples of those column sequences corresponding to
a set of $q$-cyclotomic coset representatives $\mod
\frac{q^d-1}{q-1}$. Then, under the mild restrictions on $d$ (cf.
(\ref{equ09})), it is shown that the maximum correlation magnitude
of the family is upper bounded by $(2d-1)\sqrt{q}+1$, and the
asymptotic size, as $q\rightarrow \infty$, of that is
$\frac{(M-1)q^{d-1}}{d}$. Also, we derived an exact but less
explicit expression of the size of the family of sequences by using
a result of Yucas \cite{ref12}. This generalizes the pioneering work
of Yu and Gong for $d=2$ case in \cite{ref09} and \cite{ref10}.

\appendices

\section*{Acknowledgment}
I would like to thank Prof. H.-Y. Song for drawing  the paper
\cite{ref10} to my attention. [This work was supported by National
Foundation of Korea Grant funded by the Korean
Government(2009-0072514)].

\ifCLASSOPTIONcaptionsoff
  \newpage
\fi



%

%


\begin{IEEEbiographynophoto}{Dae San Kim(M'05)}
received the B.S. and M. S. degrees in mathematics from Seoul
National University, Seoul, Korea, in 1978 and 1980, respectively,
and the Ph.D. degree in mathematics from University of Minnesota,
Minneapolis, MN, in 1989. He is a professor in the Department of
Mathematics at Sogang University, Seoul, Korea. He has been there
since 1997, following a position at Seoul Women's University. His
research interests include number theory(exponential sums, modular
forms, zeta functions) and coding theory. He has been an editor of
Journal of the Korean Mathematical Society since 2005.
\end{IEEEbiographynophoto}

\end{document}